\documentclass[aps,prc]{revtex4}
\usepackage{graphics}
\usepackage{epsfig}
\usepackage{amsmath}

\begin{document}

\title{Re-analysis of the Nucleon Space- and Time-like 
Electromagnetic Form Factors in a Two-component
Model}
\author{R. Bijker$^1$ and F. Iachello$^2$}
\affiliation{$^1$ Instituto de Ciencias Nucleares, 
Universidad Nacional Aut\'onoma de M\'exico, 
Apartado Postal 70-543, 04510 M\'{e}xico, D.F., M\'{e}xico \\
$^2$ Center for Theoretical\ Physics, Sloane Physics Laboratory, 
Yale University, New Haven, CT 06520-8120}
\date{May 11, 2004}

\begin{abstract}
Recent experimental data on space-like and time-like form factors of the
nucleon are analyzed in terms of a two-component model with a quark-like
intrinsic $q^3$ structure and $q\bar{q}$ pairs.

\

\end{abstract}

\maketitle

Recent experiments on the electromagnetic form factors of the proton \cite
{jones,rp} and the neutron \cite{rn} using the recoil polarization technique
have shown a dramatically different picture of the nucleon as compared with
a previously accepted picture \cite{walker,andivahis}. Leaving aside the
question of whether or not the experimental results are in disagreement with
each other \cite{arrington}, which is the subject of many theoretical
investigations related to the role of two-photon contributions \cite
{twophoton}, the new experiments for the proton \cite{jones,rp} are in
excellent agreement with a model of the nucleon put forward in 1973 \cite
{IJL} wherein the external photon couples both to an intrinsic structure and
to a meson cloud through the intermediate vector mesons ($\rho$, $\omega$%
, $\varphi$). On the contrary, the new experiments for the neutron \cite{rn}
are in agreement with the 1973 model up to $Q^{2}\sim 1$ (GeV/c)$^{2}$, but
not so for higher values of $Q^2$ \cite{iac1}. It is of great current
interest to understand whether a modification of the 1973 parametrization
can bring the calculation in agreement with both proton and neutron data.

We use the formalism of \cite{IJL} and introduce Dirac, $F_{1}(Q^{2})$, and
Pauli, $F_{2}(Q^{2})$, form factors. The observed Sachs form factors, $G_{E}$
and $G_{M}$ can be obtained from $F_{1}$ and $F_{2}$ by the relations 
\begin{eqnarray}
G_{M_{p}} &=&\left( F_{1}^{S}+F_{1}^{V}\right) +\left(
F_{2}^{S}+F_{2}^{V}\right) ~,  \notag \\
G_{E_{p}} &=&\left( F_{1}^{S}+F_{1}^{V}\right) -\tau \left(
F_{2}^{S}+F_{2}^{V}\right) ~,  \notag \\
G_{M_{n}} &=&\left( F_{1}^{S}-F_{1}^{V}\right) +\left(
F_{2}^{S}-F_{2}^{V}\right) ~,  \notag \\
G_{E_{n}} &=&\left( F_{1}^{S}-F_{1}^{V}\right) -\tau \left(
F_{2}^{S}-F_{2}^{V}\right) ~,  \label{ff1}
\end{eqnarray}
where we have introduced the isoscalar, $F^{S}$, and isovector, $F^{V}$,
form factors, and used $\tau =Q^{2}/4M_{N}^{2}$.

In 1973, the Dirac form factor was attributed to both the intrinsic
structure and the meson cloud, while the Pauli form factor was attributed
entirely to the meson cloud. Since this model was previous to the
development of QCD, no explicit reference was made to the nature of the
intrinsic structure. In this article, we identify the intrinsic structure
with a three valence quark structure and reanalyze the situation. In
particular, we study the question of whether or not there is a coupling to
the intrinsic structure also in the Pauli form factor $F_{2}$. Relativistic
constituent quark models in the light-front approach \cite{miller,salme} 
point to the occurrence of such a coupling. In the meantime, the
development of perturbative QCD (p-QCD) \cite{brodsky} has put some
constraints to the asymptotic behavior of the form factors, namely that the
non-spin-flip form factor $F_{1}\rightarrow 1/Q^{4}$ and the spin-flip form
factor $F_{2}\rightarrow 1/Q^{6}$. This behavior has been very recently
confirmed in a perturbative QCD\ re-analysis \cite{belitsky}. The 1973
parametrization, even if it was introduced before the development of p-QCD,
had this behavior. In modifying it, we insist on maintaining the asymptotic
behavior of p-QCD and introduce in $F_{2}^{V}$ a term of the type $%
g(Q^{2})/(1+\gamma Q^{2})$. The parametrization we use is therefore: 
\begin{eqnarray}
F_{1}^{S}(Q^{2}) &=&\frac{1}{2}g(Q^{2})\left[ 1-\beta _{\omega }-\beta
_{\varphi }+\beta _{\omega }\frac{m_{\omega }^{2}}{m_{\omega }^{2}+Q^{2}}%
+\beta _{\varphi }\frac{m_{\varphi }^{2}}{m_{\varphi }^{2}+Q^{2}}\right] ~, 
\notag \\
F_{1}^{V}(Q^{2}) &=&\frac{1}{2}g(Q^{2})\left[ 1-\beta _{\rho }+\beta _{\rho }%
\frac{m_{\rho }^{2}}{m_{\rho }^{2}+Q^{2}}\right] ~,  \notag \\
F_{2}^{S}(Q^{2}) &=&\frac{1}{2}g(Q^{2})\left[ \left( \mu _{p}+\mu
_{n}-1-\alpha _{\varphi }\right) \frac{m_{\omega }^{2}}{m_{\omega }^{2}+Q^{2}%
}+\alpha _{\varphi }\frac{m_{\varphi }^{2}}{m_{\varphi }^{2}+Q^{2}}\right] ~,
\notag \\
F_{2}^{V}(Q^{2}) &=&\frac{1}{2}g(Q^{2})\left[ \frac{(\mu _{p}-\mu
_{n}-1-\alpha _{\rho })}{1+\gamma Q^{2}}+\alpha _{\rho }\frac{m_{\rho }^{2}}{%
m_{\rho }^{2}+Q^{2}}\right] ~,  \label{ff2}
\end{eqnarray}
with $\mu _{p}=2.793$ and $\mu _{n}=-1.913$. This parametrization insures
that the three-quark contribution to the anomalous moment is purely
isovector, as given by $SU(6)$. For the intrinsic form factor we use $%
g(Q^{2})=(1+\gamma Q^{2})^{-2}$. This form is consistent with p-QCD and in
addition is the form used in our approach to the intrinsic three quark
structure by means of algebraic methods \cite{bijker}. The values of the
masses here are the standard ones: $m_{\rho }=0.776$ GeV, $m_{\omega }=0.783$
GeV, $m_{\varphi }=1.019$ GeV. The five coefficients, $\beta _{\rho }$, $%
\beta _{\omega }$, $\beta _{\varphi }$, $\alpha _{\rho }$, $\alpha _{\varphi
}$ and the value of $\gamma $ are fitted to the data. Before comparing to
the data, two modifications are needed in Eq.~(\ref{ff2}). The first
modification is crucial for the small $Q^{2}$ behavior and arises from the
large width of the $\rho $ meson. This is taken into account as in \cite{IJL}
by the replacement \cite{frazer} 
\begin{equation}
\frac{m_{\rho }^{2}}{m_{\rho }^{2}+Q^{2}}\rightarrow \frac{m_{\rho
}^{2}+8\Gamma _{\rho }m_{\pi }/\pi }{m_{\rho }^{2}+Q^{2}+\left( 4m_{\pi
}^{2}+Q^{2}\right) \Gamma _{\rho }\alpha (Q^{2})/m_{\pi }}~,  \label{ff3}
\end{equation}
with 
\begin{equation}
\alpha \left( Q^{2}\right) =\frac{2}{\pi }\left[ \frac{4m_{\pi }^{2}+Q}{Q}%
\right] ^{1/2}\ln \left( \frac{\sqrt{4m_{\pi }^{2}+Q^{2}}+\sqrt{Q^{2}}}{%
2m_{\pi }}\right) ~.  \label{ff4}
\end{equation}
Since our intent is to compare with \cite{IJL} we use the same value of the
effective width $\Gamma _{\rho }=0.112$ GeV.

The second (not important for the present range of $Q^{2}$ measurements) is
the logarithmic dependence of pertubative QCD. This can be taken into
account by the replacement 
\begin{equation}
Q^{2}\rightarrow Q^{2}\frac{\ln \left[ \left( \Lambda ^{2}+Q^{2}\right)
/\Lambda _{QCD}^{2}\right] }{\ln \left[ \Lambda ^{2}/\Lambda _{QCD}^{2}%
\right] }~,
\end{equation}
with $\Lambda =2.27$ GeV and $\Lambda _{CQD}=0.29$ GeV \cite{gari}. Since
this change gives rise to small corrections below $Q^{2}=10$ GeV$^{2}$, we
neglect it in the present paper.

\begin{figure}
\centering
\epsfig{file=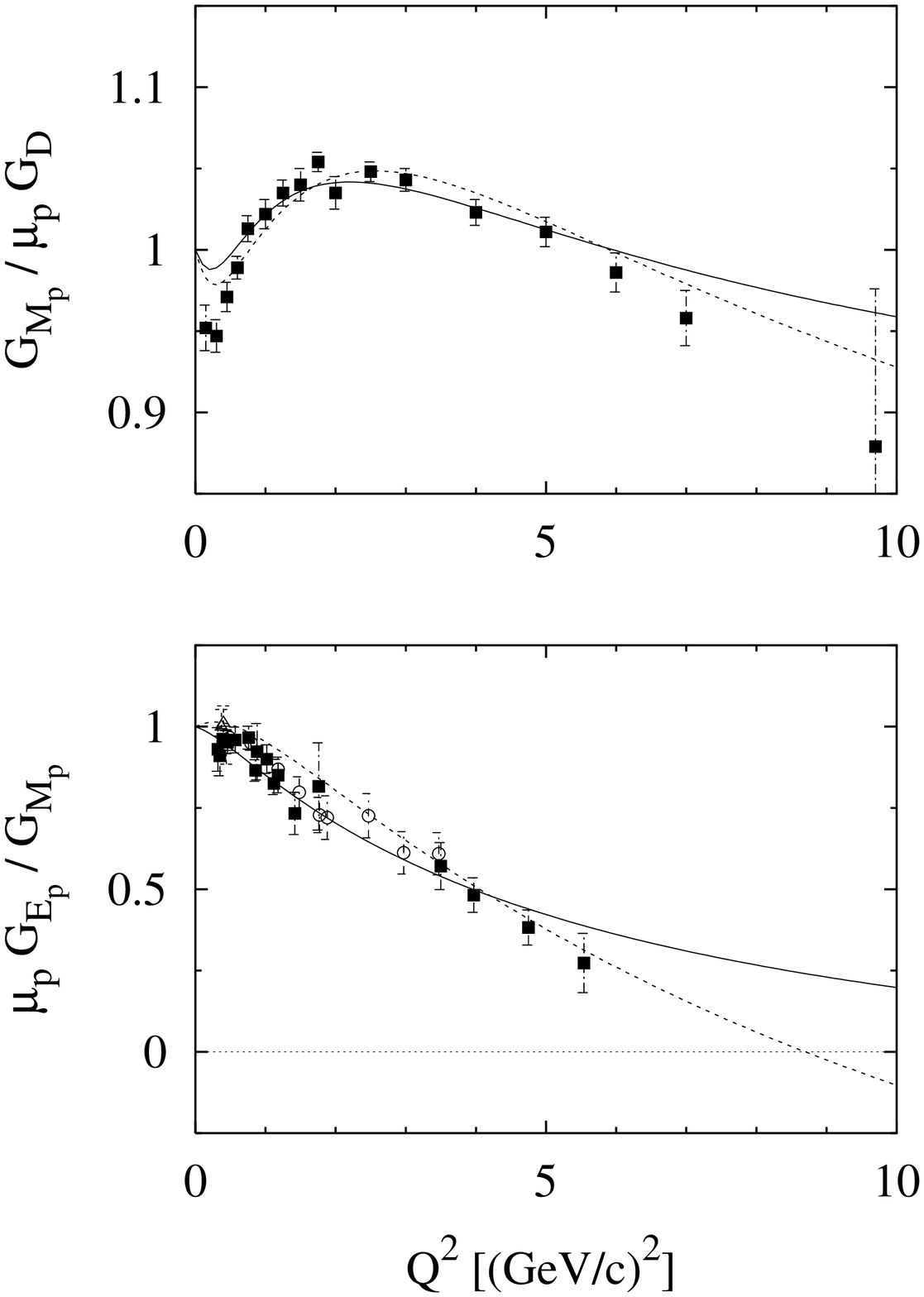,width=12cm,height=16cm}
\caption{Comparison between experimental and theoretical space-like nucleon 
form factors. Top panel: the proton magnetic form factor $G_{M_p}/\mu_p G_D$. 
The experimental values are taken from \protect\cite{walker}. 
Bottom panel: the ratio $\mu _{p}G_{E_{p}}/G_{M_{p}}$. The experimental 
data included in the fit are taken from \protect\cite{rp} (filled squares).  
Additional data, not included in the fit, are taken from 
\protect\cite{jones} (open circles) and \protect\cite{pospi} (open triangles). 
The solid lines are from the present analysis and the dashed lines from 
\protect\cite{IJL}.}
\label{pspace}
\end{figure}

\begin{figure}
\centering
\epsfig{file=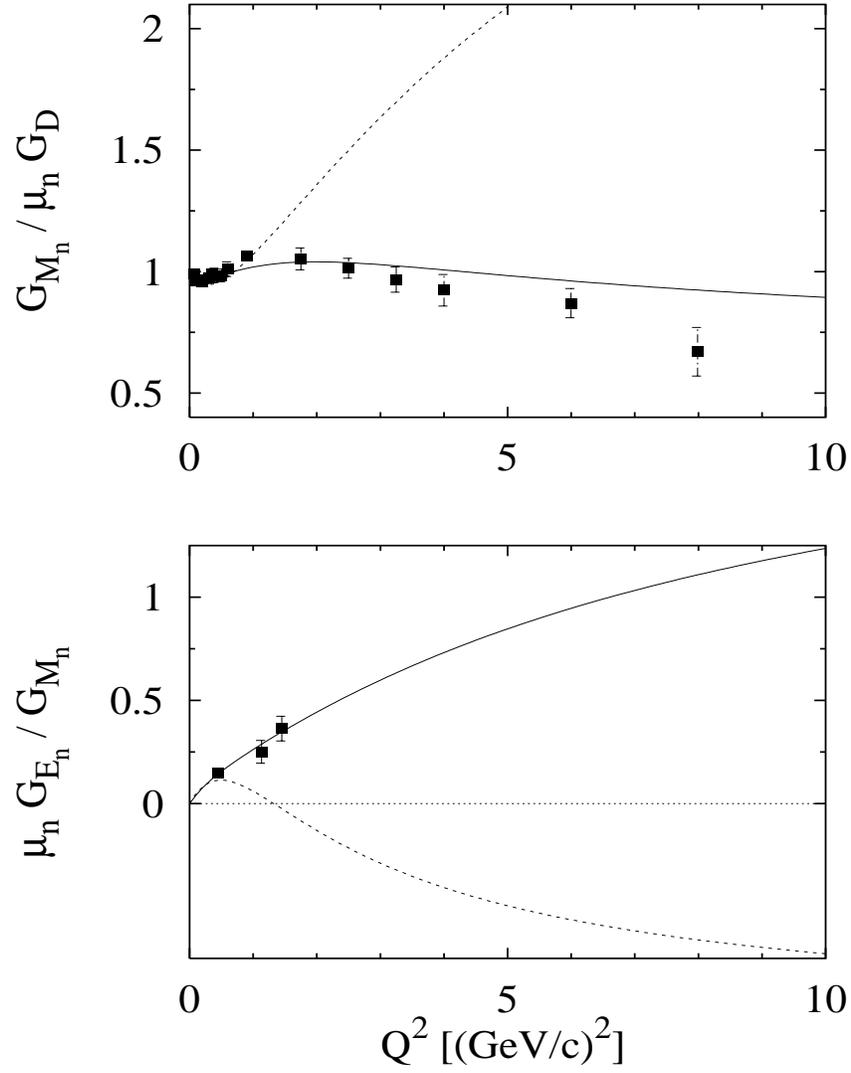,width=12cm,height=16cm}
\caption{Comparison between experimental and theoretical space-like nucleon
form factors. Top panel: the neutron magnetic form factor $G_{M_n}/\mu_n G_D$. 
The experimental data are taken from \protect\cite{gmn}. 
Bottom panel: the ratio $\mu_n G_{E_n}/G_{M_n}$. The experimental data 
are taken from \protect\cite{rn}. The solid lines are from the present 
analysis and the dashed lines from \protect\cite{IJL}.}
\label{nspace}
\end{figure}

\begin{figure}
\centering
\epsfig{file=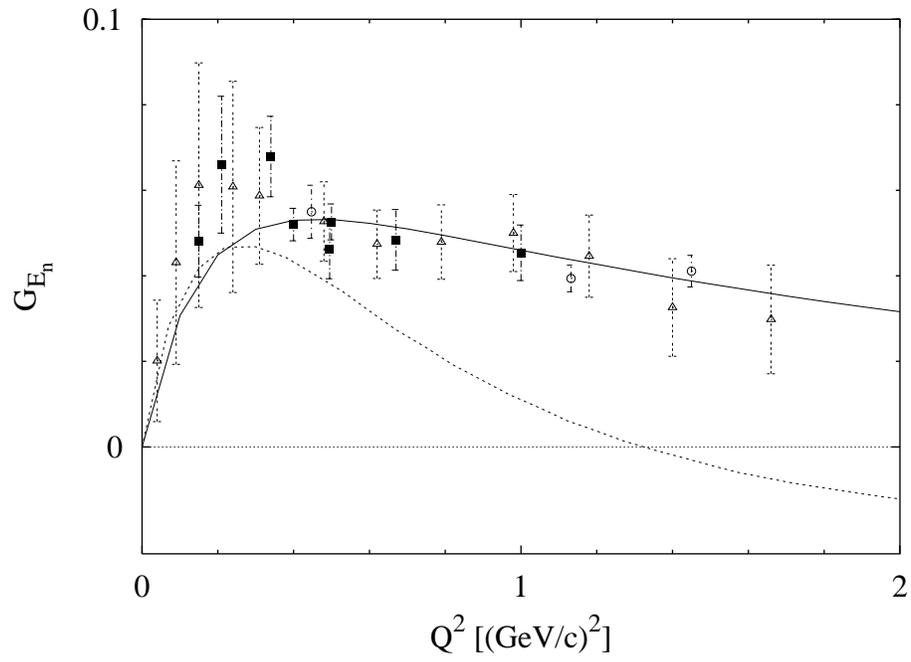}
\caption{Comparison between experimental and theoretical space-like nucleon 
form factors: the neutron electric form factor $G_{E_n}$. The experimental 
data included in the fit are taken from \protect\cite{gen} (filled squares).  
Additional data, not included in the fit, are taken from 
\protect\cite{rn} (open circles) and \protect\cite{schiavilla} 
(open triangles). The solid lines are from the present analysis and the 
dashed lines from \protect\cite{IJL}.}
\label{electric}
\end{figure}

We have determined the five coefficients $\beta _{\rho },\beta _{\omega
},\beta _{\varphi }$ and $\alpha _{\rho },\alpha _{\varphi }$ and the
parameter $\gamma $ by a fit to recent data on electromagnetic form factors.
Because of the inconsistencies between different data sets, most notably
between those obtained from recoil polarization and Rosenbluth separation,
the choice of the data to which to fit plays an important role in the final
outcome. We have used recoil polarization JLab data for the ratios 
$R_{p}=\mu_{p}G_{E_{p}}/G_{M_{p}}$ and $R_{n}=\mu _{n}G_{E_{n}}/G_{M_{n}}$ 
and Rosenbluth separation data, mostly from SLAC, for $G_{M_{p}}$ and 
$G_{M_{n}}$, as well as some recent measurements of $G_{E_n}$.  
The data actually used in the fit are quoted in the captions 
to Figs.~\ref{pspace}-\ref{electric} 
and are indicated by filled squares in those figures. The values of the
parameters that we extract are: $\beta _{\rho }=0.512$, $\beta _{\omega
}=1.129$, $\beta _{\varphi }=-0.263$, $\alpha _{\rho }=2.675$, $\alpha
_{\varphi }=-0.200$ and $\gamma =0.515$ (GeV/c)$^{-2}$. These values differ
somewhat from those obtained in the 1973 fit, although they retain most of
their properties, namely a large coupling to the $\omega $ meson in $F_{1}$
and a very large coupling to the $\rho $ meson in $F_{2}$. Also the spatial
extent of the intrinsic structure is somewhat larger than in \cite{iac1}, $%
\langle r^{2}\rangle ^{1/2}\simeq 0.49$ fm instead of $\simeq 0.34$ fm.

Fig.~\ref{pspace} shows a comparison between the calculation with parameters
given above and proton data for $R_{p}=\mu _{p}G_{E_{p}}/G_{M_{p}}$ (bottom
panel) and for $G_{M_{p}}/\mu _{p}G_{D}$, where 
$G_{D}=(1+Q^{2}/0.71)^{-2}$ (top panel). 
In this figure, the 1973 calculation, with no direct
coupling to $F_{2}^{V}$ , is also shown. One can see that the inclusion of
the direct coupling pushes the zero in $R_{p}$ to larger values of $Q^{2}$
(in \cite{iac1} the zero is at $\simeq$ 8 (GeV/c)$^{2}$). We note that any
model parametrized in terms of $F_{1}$ and $F_{2}$ will produce results for $%
R_{p}$ that are in qualitative agreement with the data, such as a soliton
model \cite{soliton} or relativistic constituent quark models 
\cite{miller,simula}. 
Perturbation expansions of relativistic effects also produce
results that go in the right direction \cite{genova}. Fig.~\ref{nspace}
shows the same comparison, but with neutron data: for $R_{n}=\mu
_{n}G_{E_{n}}/G_{M_{n}}$ (bottom panel) and for $G_{M_{n}}/\mu _{n}G_{D}$
(top panel). Contrary to the case of the 1973 parametrization, the present
parametrization is in excellent agreement with the neutron data. This is
emphasized in Fig.~\ref{electric} where the electric form factor of the
neutron is shown and compared with additional data not included in the 
fit. However, as
one can see from Fig.1, the excellent agreement with the neutron data is at
the expense of a slight disagreement with proton data. To settle the
question of consistency between proton and neutron space-like data, one
needs to: (i) Measure $\mu _{p}G_{E_{p}}/G_{M_{p}}$ beyond 6 (GeV/c)$^{2}$.
This is the approved Jlab experiment E01-109 \cite{perdrisat}. (ii) Measure $%
G_{M_{n}}$ beyond 2 (GeV/c)$^{2}$. This experiment is in the course of
analysis \cite{dejager}. (iii) Measure $G_{E_{n}}$ beyond 1.4 (GeV/c)$^{2}$.
This is the proposed experiment JLab PR04-003 \cite{madey1}.

\begin{figure}
\centering
\epsfig{file=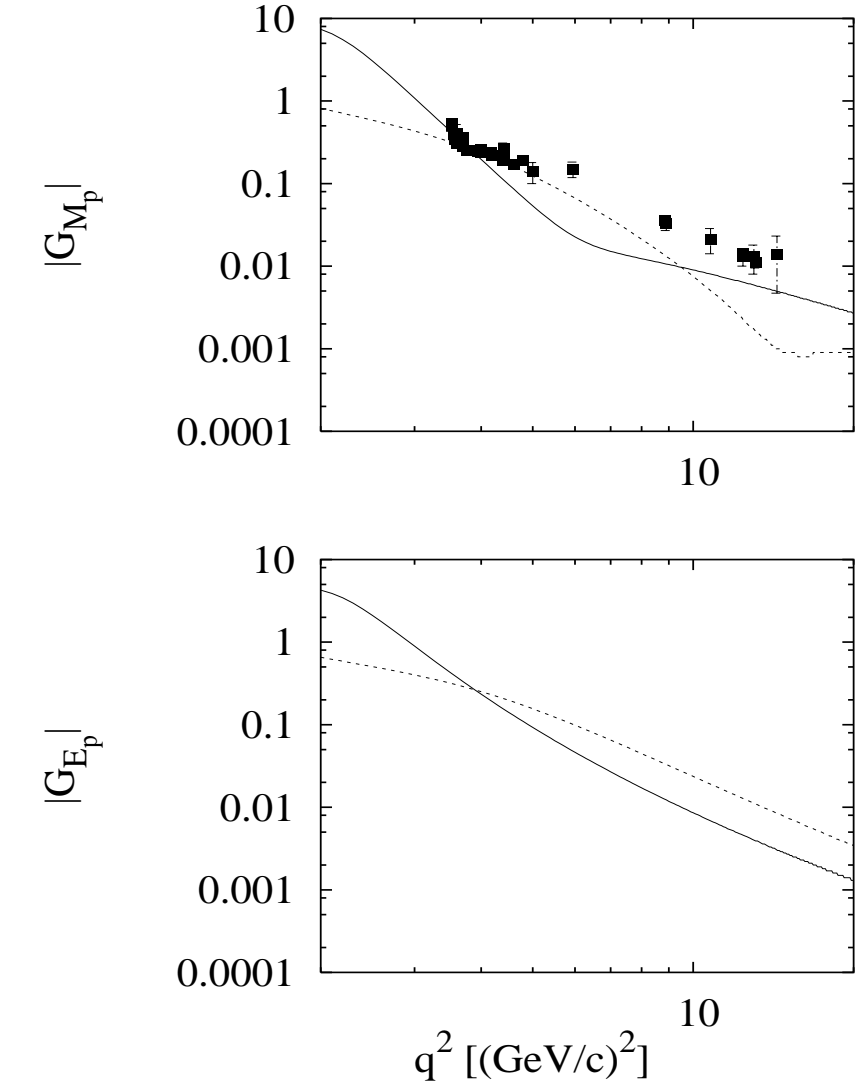,width=12cm,height=16cm}
\caption{Comparison between experimental and theoretical time-like nucleon 
form factors. Top panel: the proton magnetic form factor $|G_{M_{p}}|$. 
The experimental values are taken from \protect\cite{ptime} 
under the assumption $|G_{E_{p}}|=|G_{M_{p}}|$. 
Bottom panel: the proton electric form factor $|G_{E_{p}}|$. 
The solid lines are from the present analysis and the dashed lines 
from \protect\cite{wan}.}
\label{ptime}
\end{figure}

\begin{figure}
\centering
\epsfig{file=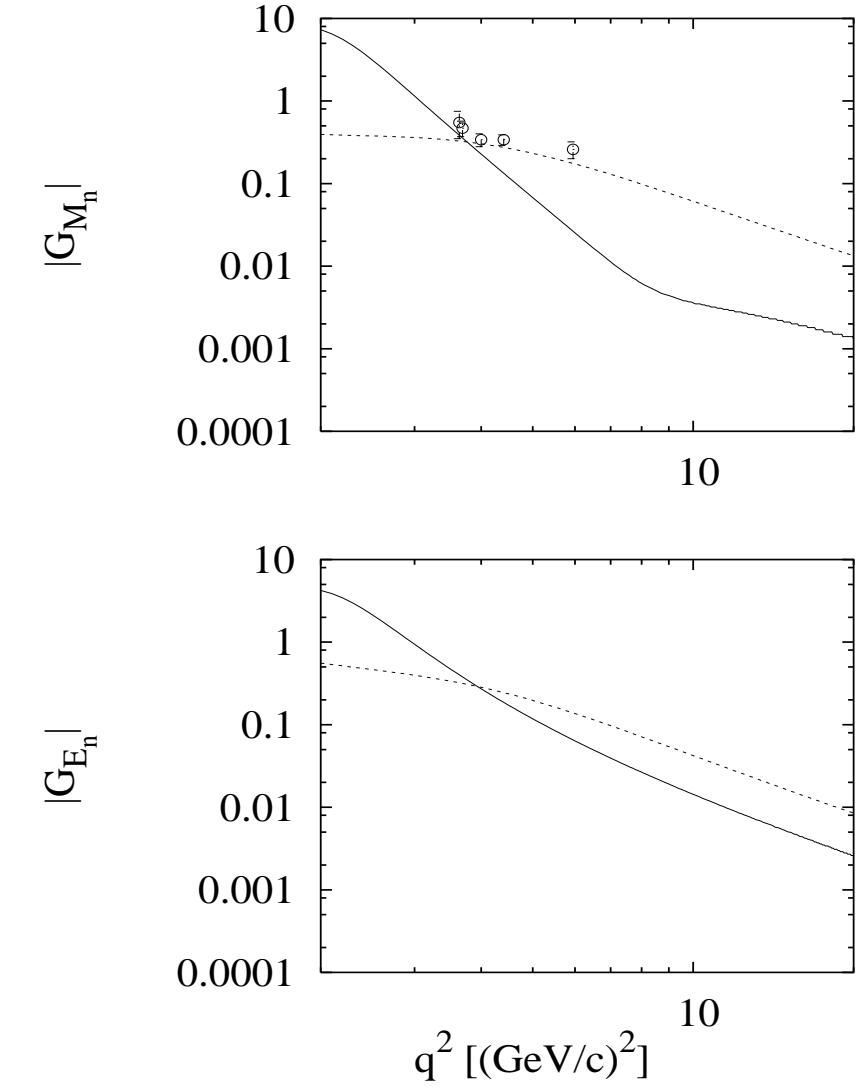,width=12cm,height=16cm}
\caption{Comparison between experimental and theoretical time-like nucleon 
form factors. Top panel: the neutron magnetic form factor $|G_{M_{n}}|$. 
The experimental values, not included in the fit, 
are taken from \protect\cite{ntime} under the assumption 
$|G_{E_{n}}|=0$.  
Bottom panel: the neutron electric form factor $|G_{E_{n}}|$. 
The solid lines are from the present analysis and the dashed lines 
from \protect\cite{wan}.}
\label{ntime}
\end{figure}

Recently it has been suggested that time-like form factors be also used in a
global understanding of the structure of the nucleon \cite{tomasi,iac21}. 
The time-like structure of the nucleon form factors within the
framework of the 1973 parametrization has been recently analyzed \cite{wan}.
We use here the same method to analyse the time-like structure of the
form-factors discussed in this note. The method consists in analytically
continuing the intrinsic structure to \cite{iac21} 
\begin{equation}
g(q^{2})=\frac{1}{(1-\gamma e^{i\theta }q^{2})^{2}}~,  \label{ff5}
\end{equation}
where $q^{2}=-Q^{2}$ and $\theta $ is a phase. The contribution of the $\rho 
$ meson is analytically continued for $q^{2}>4m_{\pi }^{2}$ as 
\cite{frazer}
\begin{equation}
\frac{m_{\rho }^{2}}{m_{\rho }^{2}-q^{2}}\rightarrow \frac{m_{\rho
}^{2}+8\Gamma _{\rho }m_{\pi }/\pi }{m_{\rho }^{2}-q^{2}+(4m_{\pi
}^{2}-q^{2})\Gamma _{\rho }[\alpha (q^{2})-i\beta (q^{2})]/m_{\pi }}~,
\label{ff6}
\end{equation}
where 
\begin{eqnarray}
\alpha (q^{2}) &=&\frac{2}{\pi }\left[ \frac{q^{2}-4m_{\pi }^{2}}{q^{2}}%
\right] ^{1/2}\ln \left( \frac{\sqrt{q^{2}-4m_{\pi }^{2}}+\sqrt{q^{2}}}{%
2m_{\pi }}\right) ~,  \notag \\
\beta (q^{2}) &=&\sqrt{\frac{q^{2}-4m_{\pi }^{2}}{q^{2}}}~.  \label{ff7}
\end{eqnarray}
Our results for time-like form factors are shown in Figs.~\ref{ptime} and~%
\ref{ntime} together with those of \cite{wan}. The phase $\theta $ obtained
from a best fit to the proton data is $\theta =0.397$ rad $\simeq
22.7^{\circ }$, again somewhat different than the value $\simeq 53^{\circ }$
obtained in \cite{wan}. It should be noted that the correction to the large $%
q^{2}$ data discussed in \cite{wan} has not been done in these figures. One
can see from these figures that while the proton form factor, $|G_{M_{p}}|$,
obtained from analytic continuation of the present parametrization is in
marginal agreement with data, the neutron form factor, $|G_{M_{n}}|$, is in
major disagreement. This result points once more to the inconsistency
between neutron space-like and time-like data already noted by Hammer 
\textit{et al.}  
\cite{meissner}, and in \cite{wan}. A remeasurement of neutron time-like data
at FRASCATI-DAFNE \cite{baldini} would help resolving this inconsistency.
The result presented here is in contrast with that of the 1973
parametrization that was in good agreement with both proton and neutron
time-like form factors. In Figs.~\ref{ptime} and~\ref{ntime}, the electric
form factors $|G_{E_{p}}|$ and $|G_{E_{n}}|$ are also shown for future use
in the extraction of $|G_{M_{p}}|$ and $|G_{M_{n}}|$ from the data. This
figure shows that the assumptions $|G_{E_{p}}|=|G_{M_{p}}|$ and $%
|G_{E_{n}}|=0$ used in the extraction of the magnetic form factors from the
experimental data, are not always justified.

In conclusion, we have performed a re-analysis of the combined space- 
and time-like data on the electromagnetic form factors of the nucleon and
found that one can obtain a good fit to the space-like neutron form factors
measured recently, but this is at the expense of a slight deterioration of
the fit for proton space-like data and especially a failure to describe
neutron time-like data. The picture emerging from the fit reported here is
that of an intrinsic structure slightly larger in spatial extent than that
of \cite{iac1}, $\langle r^{2}\rangle ^{1/2}\simeq 0.49$ fm instead of $0.34$
fm, and a contribution of the meson cloud ($q\bar{q}$ pairs) slightly
smaller in strength than that of \cite{iac1}, $\alpha _{\rho }=2.675$
instead of $3.706$.

This work was supported in part by Conacyt, Mexico, and in part by DOE Grant
No. DE-FG-02-91ER40608. We wish to thank Richard Madey for having stimulated
it and Kees DeJager for his continuing interest.

\clearpage

\end{document}